\documentclass[%
reprint,
superscriptaddress,
amsmath,amssymb,
aps,
prb,
]{revtex4-2}
\usepackage[section]{placeins}
\usepackage[dvipsnames]{xcolor}
\usepackage{algpseudocode}
\usepackage{graphicx}
\usepackage[caption=false]{subfig}
\usepackage{dcolumn}
\usepackage{array} 
\usepackage{bm}
\usepackage{tikz}
\usepackage{amsmath}
\usetikzlibrary{quantikz}
\usepackage{hyperref}
\hypersetup{
  colorlinks,
  citecolor=blue,
  linkcolor=blue,
  urlcolor=blue}
\usepackage{tikz}
\usetikzlibrary{quantikz}
\usepackage{adjustbox}

\newcommand{\beginsupplement}{
        \setcounter{table}{0}
        \renewcommand{\thetable}{S\arabic{table}}
        \setcounter{figure}{0}
        \renewcommand{\thefigure}{S\arabic{figure}}
        \setcounter{equation}{0}
        \renewcommand{\theequation}{S\arabic{equation}}
        \setcounter{page}{1} 
        \renewcommand{\thepage}{S\arabic{page}} 
        \onecolumngrid
}

\begin{document}

\preprint{APS/123-QED}
\title{Quantum Computing and Error Mitigation with Deep Learning for Frenkel Excitons}

\author{Yi-Ting Lee}
\affiliation{Department of Materials Science and Engineering, University of Illinois at Urbana-Champaign, Urbana, IL 61801, USA}

\author{Vijaya Begum-Hudde}
\affiliation{Department of Materials Science and Engineering, University of Illinois at Urbana-Champaign, Urbana, IL 61801, USA}

\author{Barbara A.\ Jones}
\thanks{manuscript submitted posthumously}
\affiliation{IBM Research Almaden Lab, 650 Harry Rd, San Jose, CA 95120}

\author{Andr\'e Schleife}
\email{schleife@illinois.edu}
\affiliation{Department of Materials Science and Engineering, University of Illinois at Urbana-Champaign, Urbana, IL 61801, USA}
\affiliation{Materials Research Laboratory, University of Illinois at Urbana-Champaign, Urbana, IL 61801, USA}
\affiliation{National Center for Supercomputing Applications, University of Illinois at Urbana-Champaign, Urbana, IL 61801, USA}

\date{\today}

\begin{abstract}
\normalsize
Quantum computers, currently in the noisy intermediate-scale quantum (NISQ) era, have started to provide scientists with a novel tool to explore quantum physics and chemistry. 
While several electronic systems have been extensively studied, Frenkel excitons, as prototypical optical excitations, remain among the less-explored applications. 
Here, we first use variational quantum deflation to calculate the eigenstates of the Frenkel Hamiltonian and evaluate the observables based on the oscillator strength for each eigenstate.
Furthermore, using NISQ quantum computers requires performing error mitigation techniques alongside simulations.
To deal with noisy qubits, we developed a deep-learning-based framework combined with a post-selection technique to learn the noise pattern and mitigate the error.
Our mitigation methods work well and outperform the conventional post-selection and remain valid on real hardware. 
\end{abstract}

\maketitle

\section{\label{sec:intro}Introduction}

In the early 1980s, the concept of a quantum computer was first proposed to solve quantum mechanical problems \cite{quantumcomputer}.
Since then, computation based on quantum computers has been actively studied, and several important physics models have been simulated, including the Ising model \cite{cervera2018exact}, the Hubbard model \cite{cade2020strategies, cai2020resource, stanisic2022observing}, and the Heisenberg Hamiltonian \cite{cubitt2018universal, tacchino2020quantum}.
The inherent properties of qubits, such as superposition and entanglement, facilitate the simulation of complex interacting many-body systems \cite{feynman2018simulating}.
Furthermore, the possible advantage of quantum computers in storing and processing information might render them more efficient than classical computers also in the realms of machine learning \cite{biamonte2017quantum} and Monte Carlo simulations \cite{montanaro2015quantum}.
Quantum computers have found applications in investigating high-energy physics \cite{nachman2021quantum}, open quantum systems \cite{wang2011quantum, hu2020quantum}, portfolio optimization \cite{orus2019quantum}, and various other domains \cite{rubin2024quantum}.

Over the last decade, early stage quantum computing hardware has been explored for solving problems in quantum chemistry and physics.
While ground-state preparation is generally considered a hard problem and may require exponential resources on quantum computers in the worst case \cite{kempe2006complexity}, the many-body wave function of an $N$ particle system can be  prepared on a quantum computer with $\mathcal{O}(N)$ effort, compared to the $\mathcal{O}(\exp{N})$ effort on a classical computer \cite{cao2019quantum}.
Hence, simulating interacting electrons on a quantum computer holds promise:
The ground-state energies of several simple molecules, including H$_2$, LiH, and BeH$_2$ \cite{kandala2017hardware} have been solved using the variational quantum eigensolver (VQE).
Additionally, several subspace methods \cite{motta2024subspace} have been proposed to compute excited states of quantum-mechanical Hamiltonians.

One particular limitation of noisy intermediate-scale quantum (NISQ) computers, is the need for error mitigation to deal with noise originating from low fault-tolerance qubits, significantly degrading the accuracy of simulation results \cite{preskill2018quantum}.
Zero-noise extrapolation (ZNE) \cite{giurgica2020digital, he2020zero} and probabilistic error cancellation (PEC) \cite{van2023probabilistic} are two commonly used mitigation techniques.
When performing ZNE, the noise is scaled up and the data is extrapolated to estimate the noiseless result. 
However, if the circuit suffers from large noise, ZNE may not provide reliable data for extrapolation \cite{giurgica2020digital}.
On the other hand, challenges for PEC include the cost of sampling the circuit, which scales exponentially with the number of qubits used \cite{temme2017error}. 

Most physics or chemistry examples studied on quantum computers have focussed on ground and excited states of electronic systems
However, excitons as two-particle electron-hole excitations are explored less.
Two exciton models are the Frenkel Hamiltonian \cite{frenkel1931transformation}, with binding energies on the order of 1 eV e.g.\ in alkali molecules \cite{rittner1951binding}, and the Wannier-Mott Hamiltonian in solids \cite{wannier1937structure}, with binding energies on the order of tens of meV.
Since the Wannier-Mott exciton is delocalized, its description usually requires careful convergence of reciprocal space sampling, leading to large rank exciton Hamiltonians \cite{fuchs2008efficient,Zhang:2021,Schleife:2018}.
Hence, in this study we focus on the highly localized Frenkel exciton that allows a description with small rank Hamiltonians, which is more amenable to treatment on current quantum computers.
To describe Frenkel excitons, the Frenkel-Davydov (FD) model \cite{davydov1964theory} is usually employed.
In this work we study this Hamiltonian on quantum computers to enhance our understanding of the capabilities and limitations of current quantum hardware.
In particular, establishing a suitable error mitigation framework for the FD Hamiltonian becomes imperative to achieve highly accurate results with real quantum hardware. 
Besides, it can offer valuable insight into error mitigation that is useful for studying other systems.

In this work, anthracene, comprising of five molecules, serves as a toy model for which we investigate the solution of the Frenkel-Davydov exciton Hamiltonian on a quantum computer, as a case study for two-particle optical excitations.
For this system, there is experimental data to compare our results to.
First, we compute all eigenvalues using the variational quantum deflation technique and then calculate the associated observables. 
To develop an error mitigation technique, we utilize the noise pattern from a real quantum device and develop a deep-learning based technique. 
Our method reduces the error of the Davydov splitting to less than 10 cm$^{-1}$ in observable measurements on a noisy simulator and remains effective on real quantum hardware.
These findings hold promise for advancing excited-state simulations and introduce new perspectives on machine-learning based noise mitigation techniques within the quantum computing community.

\section{\label{sec:method}Theory and Methods}

\subsection{Frenkel-Davydov Hamiltonian}

The Frenkel-Davydov (FD) model  \cite{frenkel1931transformation,davydov1964theory} describes a tightly bound exciton as it is found e.g.\ in organic solids composed of aromatic molecules \cite{frenkel1931transformation, davydov1964theory}.
It can be expressed in second quantization as
\begin{equation}
\label{eq:hamiltonian}
H = \sum \Omega_m B^{\dagger}_m B_m + \frac{1}{2} \sum_{m,n} V_{mn} (B^{\dagger}_m B_n + h.c.)
\end{equation}
where $\Omega_m$ represents the transition energy from ground to excited state of molecule $m$ and
$V_{mn}$ denotes the electronic coupling term between molecules $m$ and $n$.
Additionally, $B_m^{(\dagger)}$ represents the excitonic annihilation (creation) operator, and \textit{h.c.}\ stands for the Hermitian conjugate.
Solving the FD Hamiltonian determines the excitation levels of a system with excitonic effects, which is crucial for studying optical properties \cite{spano2010spectral}. 

Electronic coupling $V_{mn}$ between two molecules $m$ and $n$ is evaluated by two-body Coulomb and exchange integrals \cite{scholes1996configuration,harcourt1994rate}.
These are given as
\begin{equation}
V_{mn} = \sum_{iajb} C^m_{ia} C^n_{jb}[2(\psi^m_i \psi^m_a | \psi^n_j \psi^n_b) - (\psi^m_i \psi^n_j | \psi^n_a \psi^m_b) ],
\label{Vmn}
\end{equation}
where $\psi$ represents molecular orbitals, and the indices $i, j$ $(a, b)$ denote occupied (unoccupied) orbitals.
The parenthesis $(ab|cd)$ refers to the two-body integral in chemistry notation, expressed as $\int \psi_a(x_1)^{\ast} \psi_b(x_1) \frac{1}{r_{12}} \psi_c(x_2)^{\ast}\psi_d(x_2) dx_1dx_2$.
This Coulomb term can be calculated through integration, and can also be evaluated using the transition density cube method \cite{krueger1998calculation}, distributed transition dipole approximation \cite{sumi1999theory}, point-dipole approximation, and other efficient approaches \cite{plotz2014new}.
Since molecular orbitals from different molecules exhibit minimal overlap, exchange interactions are typically neglected, focusing solely on the contribution of the first term in Eq.\ \eqref{Vmn}.
The coefficient $C_{ia}^m$ can be interpreted as constructing singly excited determinants in the configuration interaction expansion, transitioning from orbital $i$ to orbital $a$ in molecule $m$.
These coefficients can be obtained through configuration-interaction singles calculations or time-dependent density functional theory calculations under the Tamm–Dancoff approximation \cite{hirata1999time}.

To solve the eigenvalue problem of the FD Hamiltonian, Eq.\ \eqref{eq:hamiltonian}, the wavefunction is described by an exciton basis.
In the Heitler-London approximation it is composed only of single excitations \cite{knoester2002optical} and in second quantization reads
\begin{equation}
\label{basis}
|\psi \rangle = \sum_m C_m B^{\dagger}_m |G\rangle ,
\end{equation}
where $C_m$ is the expansion coefficient, $B^{\dagger}_m$ is the localized exciton creation operator, and $|G\rangle$ represents the ground state without any exciton.
In this work, the exciton refers to the electron-hole pair formed between the highest occupied molecular orbital (HOMO) and the lowest unoccupied molecular orbital (LUMO).
This excitation is optically allowed in this work and we consider only the case of one exciton.

\begin{figure}
\includegraphics[width=1\columnwidth]{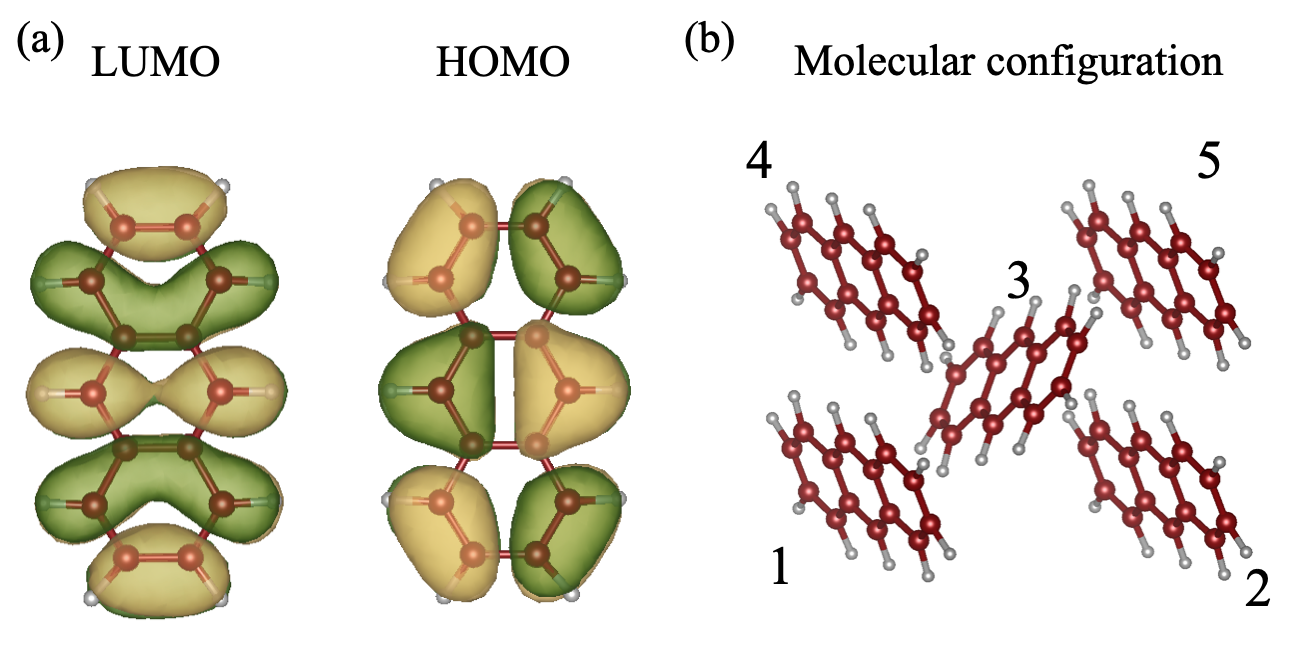}
\caption{\label{fig:config_trans}
(a) Real wave function of the HOMO and LUMO of a single anthracene molecule are shown as isosurfaces (green corresponds to positive and yellow to negative values), where the isosurfaces correspond to 95\,\% of electrons.
(b) Five anthracene molecules arranged as a 3D anthracene crystal used in our simulations, based on the experimental structure \cite{mason1964crystallography}.
Each molecule is labeled from index 1 to 5, where red and grey atoms represent carbon and hydrogen atoms, respectively. 
The interaction among three different types of nearest pairs, $V_{12}$, $V_{13}$, $V_{14}$, is taken into account in the Hamiltonian, Eq.\ \eqref{eq:hamiltonian}. 
There are two equivalent $V_{12}$, two $V_{14}$, and four $V_{13}$ interactions in this configuration.
}
\end{figure}

To study optical excitations, Davydov splitting is a key observable, as it describes the experimentally measurable energy spacing between allowed transitions.
This work studies a single layer of anthracene, approximated by five molecules as a minimum toy model, see Fig.\ \ref{fig:config_trans}.
The experimental structure, determined by X-ray diffraction \cite{mason1964crystallography}, is utilized.
Notably, the excitation energies $\Omega_m$ ($m=1, ..., 5$ in this work) are set to zero, since the diagonal elements of the Hamiltonian only cause a global shift for each eigenvalue.
The optical oscillator strength for the transition from the ground state to the $i$-th excited state is $f_i = \frac{2}{3}E_i \left(\sum_{m=1}^N C_m^i \mu_m\right)^2$ \cite{nematiaram2021bright}, where $C_m^i$ is the expansion coefficient of the $i$-th eigenstate in the exciton basis, $\mu_m$ is the transition dipole moment of the $m$-th molecule, and $E_i$ is the excitation energy.
Experimentally, the Davydov splitting of the spectral peak can be observed through optical measurements and will be used for comparison with the calculated results.

First-principles simulations of Davydov splitting on a classical computer serve as a reference throughout this work.
We first construct the FD Hamiltonian, with molecular orbitals for individual anthracene molecules, evaluated using the 6-31g** basis set and the B3LYP functional \cite{becke1992density,stephens1994ab} via the PySCF code \cite{sun2018pyscf}.
By diagonalizing the Hamiltonian, the eigenstates and eigenvalues are obtained, and the optical oscillator strengths  are evaluated using the transition dipole moment $\mu_m$ calculated via PySCF.

\subsection{Quantum Computing}

To compute the ground state of a given Hamiltonian on a quantum computer, the variational quantum eigensolver (VQE) is one of the most widely used algorithms \cite{kandala2017hardware}.
In this quantum-classical hybrid algorithm, the wave function is first described by a parameterized quantum circuit $|\psi(\lambda)\rangle$, which is constructed by applying a unitary operator $U(\lambda)$ to the initial state $|\psi\rangle$ \cite{kandala2017hardware}. 
Then the circuit is measured, followed by an evaluation of the expectation value $\langle \psi(\lambda)|H|\psi(\lambda)\rangle$. 
The parameters in the quantum circuit are optimized iteratively by a classical optimizer until the expectation value converges to the minimum.
Utilizing the variational principle, the ground state energy serves as the lower bound for this search process.
However, this method is tailored for finding the lowest eigenvalue, which does not suffice for determining the Davydov splitting, since that requires the solution also for higher eigenstates.

To compute excited states of the Frenkel-Davydov Hamiltonian, we use the variational quantum deflation (VQD) technique \cite{higgott2019variational}.
In this algorithm, the Hamiltonian for finding the $k$-th eigenstate can be constructed by adding a penalty term to the original Hamiltonian
\begin{equation}
H_k = H + \sum_{i=0}^{k-1} w_i |i\rangle \langle i|,
\end{equation}
where $|i\rangle$ represents the $i$-th eigenstate and $w_i$ needs to be sufficiently large compared to the energy scale of the original Hamiltonian \cite{higgott2019variational}. 
In this work, we choose $w_i$ larger than the ground-state energy.
VQD is also known as orthogonality-constrained VQE, as it assumes that the overlap between different eigenstates is zero and allows all eigenstates to share the same ansatz with different parameters.
To implement this algorithm, a cost function for finding the eigenstate up to the $k$-th eigenstate is introduced as
\begin{equation}
\label{cost}
C(\lambda)_k= \langle \psi(\lambda)_k|H|\psi(\lambda)_k\rangle + \sum_{i=0}^{k-1} w_i | \langle\psi(\lambda)_k| \psi_i\rangle|^2.
\end{equation}
The first term is the expectation value of the Hamiltonian and the second term represents the overlap between the $k$-th eigenstate and the $i$-th eigenstate. 
To measure the overlap term, SWAP-based tests \cite{garcia2013swap} are typically conducted on quantum hardware, which largely increases the cost due to building additional quantum circuits.
In this work, instead of performing a SWAP-based test, the overlap integral is efficiently evaluated based on the properties of our proposed ansatz, as we will elaborate in the next paragraph.

\begin{figure}
\includegraphics[width=0.9\columnwidth]{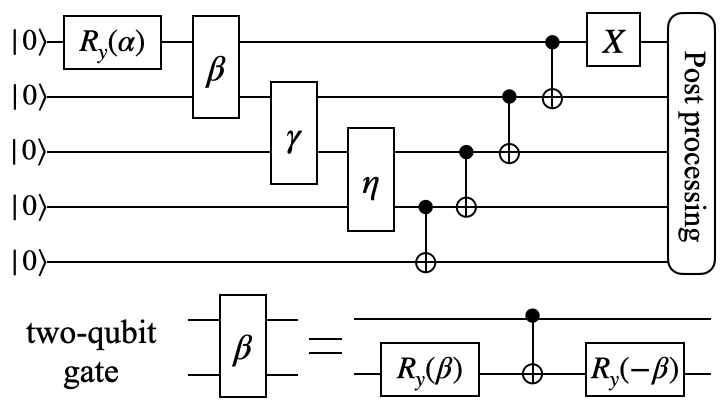}
\caption{\label{w_state_vqd} 
This circuit represents an Ansatz to create the local and entangled basis of Frenkel excitons out of the initial $|00000\rangle$ ground state.
The X gate flips 0 and 1.
The two-qubit controlled-rotation gates, labeled by Greek letters, comprise of two rotation gates, $R_y$, around the $y$ axis with opposite angles and one CNOT gate.
Post-processing gates can include post-rotation gates, post-selection gates, or other quantum gates, depending on the post-processing strategies.
This circuit can be extended with respect to the number of qubits and, thus, the rank of the Frenkel Hamiltonian.
}
\end{figure}

The wave function ansatz for excitons, Eqn.~\eqref{basis}, is represented by the circuit shown in Fig.\ \ref{w_state_vqd}.
For $n$ molecules, the wave function can be written using a $n$ one-particle basis as
\begin{equation}
\label{binarypsi}
|\psi\rangle = \sum_{i=0}^{n-1} a_i |(2^{i})_2\rangle
= a_0|00...01\rangle +
...+ a_{n-1}|10...00\rangle,
\end{equation}
where $a_i$ is an expansion coefficient. 
The total wave function has a form similar to that of the typical W-state \cite{dur2000three}, representing a superposition where only one qubit is in the '1' state while the others are in the '0' state.

To construct this ansatz, first one $R_y$ gate and several two-qubit gates are used to create $n$ separate states (see Fig.\ \ref{w_state_vqd}).
Next, sequential CNOT and $X$ gates are applied to adjust the bitstring so that only one qubit is in the '1' state, while the others are in '0'.
In this ansatz, there is only one CNOT gate for each two-qubit rotation gate, and the total number of CNOT gates for an $n$-qubit system is only $2n-3$, which is less than $3n-3$ reported in Ref.\ \cite{cruz2019efficient}.
The reduced number of CNOT gates could alleviate noise effects while still maintaining the representation of the entire phase space.
Moreover, a deterministic relationship exists between the ansatz parameters and the coefficients of the basis as shown in Appendix \ref{appendix_1}, enabling the evaluation of the overlap term in VQD without relying on SWAP-based measurements. 
This approach, developed in this work, significantly reduces noise and accelerates the VQD process.

To perform VQD simulations, we compute expectation values of the Hamiltonian by measuring quantum circuits, while the eigenstate overlap term is evaluated classically.
The Jordan-Wigner transformation \cite{jordan1993paulische} is used to map the Frenkel-Davydov (FD) Hamiltonian to qubit operators.
To measure the Hamiltonian in the $Z$-basis, which is the standard computational basis  \cite{mermin2007quantum}, we convert the $X_pX_q+Y_pY_q$ qubit operator, originating from the off-diagonal elements of the FD Hamiltonian, into the $I_pZ_q-Z_pI_q$ operator by implementing a unitary $XX+YY$ quantum gate
\begin{equation}
\label{ps_gate}
U_{XX+YY} = 
\begin{quantikz}[scale=0.9, baseline=(current bounding box.center)]
    & \gate{Rz(-\pi/4)} & \gate[2]{\sqrt{iSWAP}} & \qw \\ 
    & \gate{Rz(\pi/4)}  &                        & \qw 
\end{quantikz}
\end{equation}
This unitary gate can transform the off-diagonal elements into the $Z$-basis, as described in Ref. \cite{google2020hartree}.
The circuits are measured 8192 times, commonly referred to as shots, to sample the distribution of basis states from wavefunctions and the parameters within are optimized iteratively using the COBYLA optimizer \cite{powell2007view, conn1997convergence}.

\subsection{\label{sec:fnntrain}Error mitigation}

The VQD results from our quantum simulations show errors of the expectation values of more than 30\,\%, requiring the use of error mitigation.
In this work, we compare post selection and a deep learning technique.
Post selection reduces non-physical solutions by enforcing particle number conservation and relies on the fact that our basis contains only one exciton as shown in Eq.\ \eqref{binarypsi}.
Measured results with a multi-exciton basis, where multiple '1's appear in the binary string, will be treated as noise and discarded. 
The post-processed results will then be normalized.
It is important to note that post selection is an approximate method for obtaining ideal probabilities, and the discrepancy between the post-processed and ideal probabilities will increase as the basis size grows. 

Capturing the underlying pattern between noisy and ideal distributions is crucial for recovering the ideal distribution from the noisy one.
For this, the deep learning technique predicts noise-mitigated wavefunctions from noisy ones caused by bit-flip errors, dephasing, damping, two-qubit depolarization, and other sources \cite{martinis2003decoherence, bertet2005dephasing}, while also enforcing particle number conservation.

We use a feed-forward neural network (FNN) for deep learning the connection between noisy and noiseless data.
In this research, we construct a FNN with three hidden layers.
For training the FNN, input data is collected by sampling the noisy circuit, while the ideal distribution, which can be effectively calculated via Appendix \ref{appendix_1}, serves as labeled data. 
To account for the noise from the $XX+YY$ gates, we concatenated the ansatz with two CNOT gates (see Fig.\ \ref{DL}a), which are the primary source of error in the $\sqrt{iSWAP}$ gate, see Eq.\ \eqref{ps_gate}.
Ten percent of the overall dataset is used to benchmark the performance of the FNN, and we gradually increase the size of training/validation dataset with an 80/20 split. 
The hyperparameters in our FNN are further optimized by Bayesian optimization.
We utilize the Adam optimizer for gradient descent \cite{kingma2014adam}, with the mean absolute error (MAE) serving as the error metric. 
The training process is performed using Tensorflow 2.10.0 \cite{tensorflow2015-whitepaper} under Python 3.10.10.

We test our error mitigation framework using simulated noise based on the model of the $ibmq\_guadalupe$ device, with the corresponding parameters summarized in Table \ref{tab:hardware_parameters}, as well as real hardware noise.
After examining our technique, we perform our simulation, as well as the deep-learning-based mitigation technique, using the $ibmq\_jakarta$ quantum computer. All simulations are run using the IBM Qiskit package \cite{Qiskit}.

\section{Results and discussion}

\subsection{Hamiltonian diagonalization}

\begin{table}[h]
\begin{ruledtabular}
\begin{tabular}{ccccc}
\textrm{Excitation} & Eigenvalue (meV) & $f_i$ & transition\\
\hline
$1^\mathrm{st}$ & $-32.562$ & 0   & forbidden  \\
$2^\mathrm{nd}$ & $-24.449$ & 0.823   & allowed  \\
$3^\mathrm{rd}$ &  2.577 & 0.835  & allowed \\
$4^\mathrm{th}$ & 21.872 & 0    & forbidden \\
$5^\mathrm{th}$ & 32.562 & 0   & forbidden  \\
\end{tabular}
\end{ruledtabular}
\caption{\label{tab:classical}
Eigenvalues and oscillator strengths for each excitation in the five anthracene molecules aggregate, which are calculated classically with the FD Hamiltonian.
Note that only the second and third excitations are allowed.}
\end{table}

We first calculate the couplings, Eq.\ \eqref{Vmn}, between the three nearest-neighbor molecules (see Fig.\ \ref{fig:config_trans}), to set up the Hamiltonian, Eq.\ \eqref{eq:hamiltonian}.
These values are $V_{12}$=5.345 meV, $V_{13}$=3.969 meV, and $V_{14}$=$-27.217$ meV, respectively, and are consistent with previous theoretical work \cite{giannini2022exciton}.
After constructing the FD Hamiltonian, we compute the eigenvalues and the oscillator strength for each excitation, as summarized in Table \ref{tab:classical}. 
The data in Table \ref{tab:classical} indicates that transitions for the second and third excitations will be allowed, resulting in Davydov splitting in the optical spectrum. 
The corresponding energy splitting is 27.006 meV, which is 218.75 cm$^{-1}$, consistent with the reported Davydov splitting of 190 to 220 cm$^{-1}$ \cite{clark1970anisotropy,philpott1971calculation,mahan1964davydov}.
We later will compare these simulated eigenvalues and Davydov splittings with data computed on quantum hardware.

Finally, we validate our variational quantum deflation (VQD) algorithm on a noiseless quantum simulator, by ensuring that the eigenvalues of the Hamiltonian, Eq.\ \eqref{eq:hamiltonian} match the exact classical result.
The simulated results are shown in Fig.\ \ref{noiseless} and we find good agreement.

\subsection{Error mitigation: Post selection}

\begin{figure}[h]
\includegraphics[width=0.99\columnwidth]{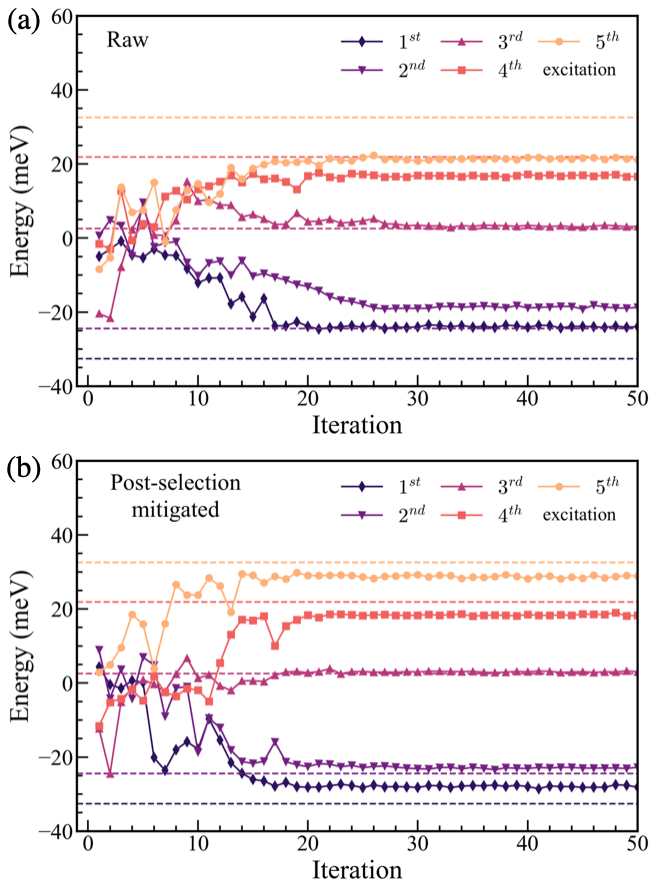}
\caption{\label{ps}Quantum simulations of Frenkel Hamiltonian with a quantum simulator with the noise model from $ibmq\_guadalupe$ device. 
(a) Raw results without performing any error mitigation. 
(b) Post-selection mitigated results. 
Colored lines are the results under noisy environment and dashed line are the exact ground truth.
}
\end{figure}

\begin{table}[h]
\begin{ruledtabular}
\begin{tabular}{cccc}
\textrm{Excitation} & \textrm{Exact (meV)}& \textrm{Raw (meV)}& \textrm{PS (meV)} \\
\hline
$1^{st}$ &  $-32.562$ & $-23.683$ (8.879)& $-27.893$ (4.669)  \\
$2^{nd}$ & $-24.449$ & $-18.751$ (5.698)& $-22.743$ (1.706) \\
$3^{rd}$ & 2.577 &2.999 (0.422) & 2.928 (0.351)\\
$4^{th}$ & 21.872 &16.650 (5.222)& 18.902 (2.970) \\
$5^{th}$  & 32.562 & 21.689 (10.873)& 28.693 (3.869) \\
\hline 
DS (cm$^{-1}$) & 218.75 & 176.18 (42.57) & 207.94 (10.81) \\
\end{tabular}
\end{ruledtabular}
\caption{\label{tab:error}
Comparison between the exact excitation energies and those calculated by noisy simulation without post-selection (raw) and with post-selection (PS) is presented. These energies represent the converged values from Fig. \ref{ps}, with the corresponding absolute errors shown in parentheses. DS refers to the Davydov splitting, which is the energy difference between the $2^{nd}$ and $3^{rd}$ excitations.}
\end{table}

To benchmark different error mitigation techniques and to develop our deep-learning based framework, we utilize the noise model from the $ibmq\_guadalupe$ device in the Qiskit simulator.
VQD results with simulated noise deviate from the eigenvalues calculated from classical diagonalization, as shown in Fig.\ \ref{ps}(a), and the corresponding errors are summarized in Table \ref{tab:error}.
Notably, the calculated energy is overestimated for negative values and underestimated for positive values, causing a symmetric deviation toward zero energy. 
The corresponding Davydov splitting for the noisy results is 176.18 cm$^{-1}$ with a 42.57 cm$^{-1}$ underestimation compared to the 218.75 cm$^{-1}$ evaluated analytically.

Error mitigation through post-selection reduces the errors, as summarized in Table \ref{tab:error}.
The calculated energies are closer to the classical results compared to the raw results, as shown in Fig. \ref{ps}(b).
Post selection and subsequent renormalization discards nonphysical measurements of our one-particle wave function, hence, bringing the result closer to the ideal wave function, as discussed below.
Although post-selection is one of the most efficient methods for mitigating errors, it is not ideal, as it does not capture the relationship between noiseless and noisy wave functions.
This limitation results in the error after post selection ranging from 0.351 meV to 4.669 meV for the different states (see Table \ref{tab:error}), with a Davydov splitting of 207.94 cm$^{-1}$.

\subsection{Error mitigation: Deep-learning approach}

\begin{figure*}
\includegraphics[width=0.97\textwidth]{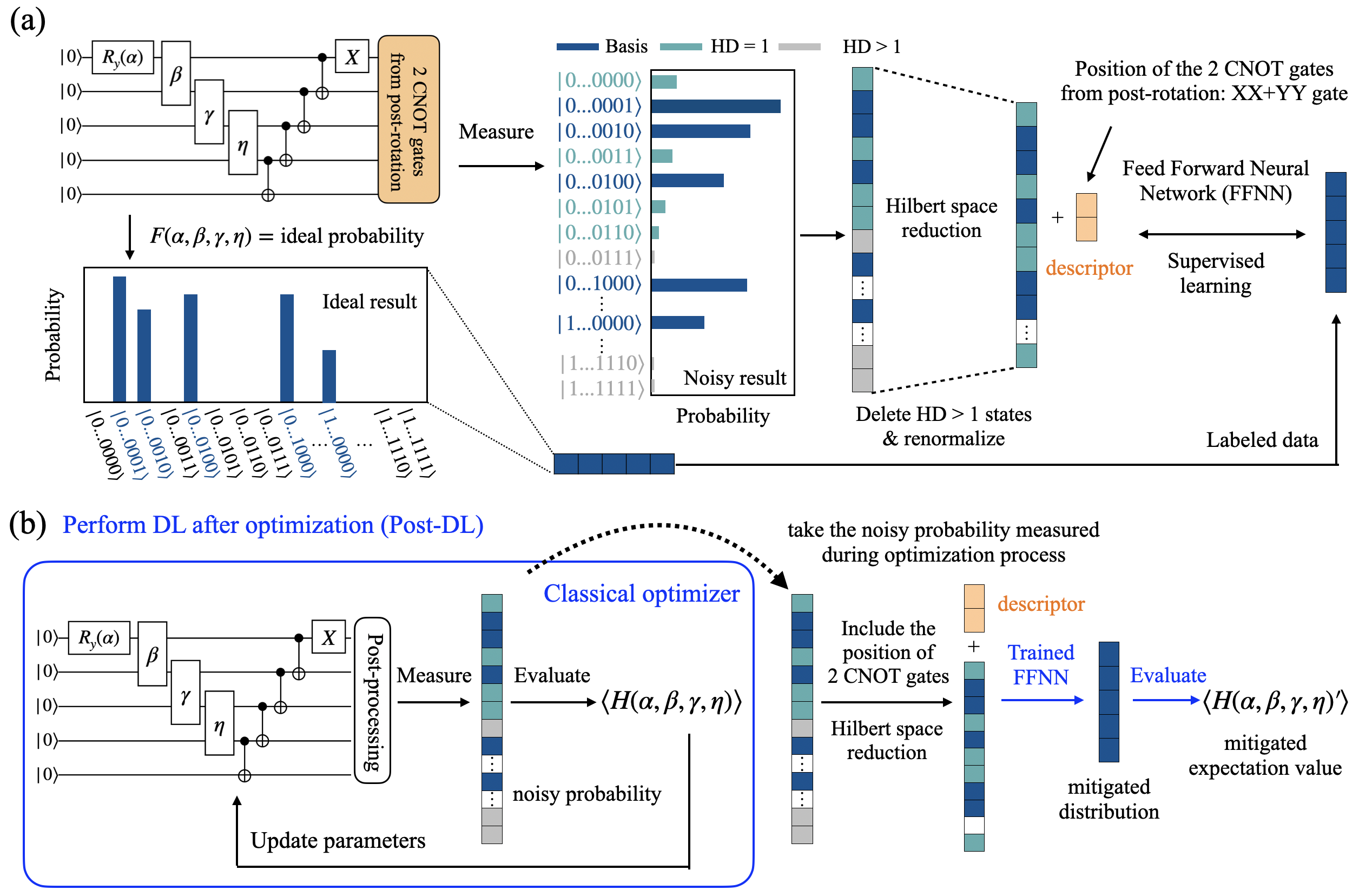}
\caption{\label{DL} Strategy for post-selection-based deep-learning mitigation technique:
(a) Process of collecting data and training feedforward neural networks. 
The original circuit concatenated with 2 CNOT gates is used to describe the noisy pattern of the entire circuit with post-selection gates. 
Random parameters are assigned in the circuit, and the circuit is measured to obtain a noisy probability distribution, which dimension will be further reduced by selecting the states within a Hamming distance (HD) equal to 1. 
For the training section, the reconstructed probability is concatenated with the information of the position of the CNOT gate as the input, and the labeled data for supervised learning will be calculated classically through the deterministic function derived in Appendix.\ref{appendix_1}. 
(b) Workflow of using deep learning as a post-processing tool to mitigate the optimized results.
Colored and dashed lines represent simulated results and exact ground truth.
}
\end{figure*}

To further improve error mitigation, we trained a feedforward neural network (FNN) on the noise pattern, following the strategy shown in Fig.\ \ref{DL}(a).
We first perform a learning curve analysis (see Fig.\ \ref{dataset_training}) using a noisy simulator and determine that one thousand datasets for training our FNN are sufficient to predict each measured circuit.
The resulting root mean square error (RMSE) of the testing data converges within $10^{-3}$.
During the training, the rectified linear unit activation function is employed, as it outperforms the sigmoid function (see Fig.\ \ref{dataset_training}).
The hyperparameters are fine-tuned through Bayesian optimization and are summarized in Table \ref{tab:bayisen}.

To address the exponential increase of the size of the Hilbert space, we compare to dimensional reduction based on Hamming distance (HD).
Specifically, we only keep the basis within an HD of 1 from the single-particle basis for training the FNN as shown in Fig.\ \ref{DL}(a), which keeps 98\,\% of the noisy wave function.
As a result, the FNN trained with HD-reduced data achieves a similar RMSE to the model trained with the raw data (see Fig.\ \ref{dataset_training}).
Reducing the dimensionality of the training dataset does not compromise the effectiveness of the FNN while ensuring the size of the wave functions scales linearly.

\begin{table}[h]
\begin{ruledtabular}
\begin{tabular}{cccc}
\textrm{Excitation} & \textrm{Exact (meV)}& \textrm{Post-DL (meV)} & \textrm{DL-VQD (meV)} \\ \hline
\rule{0pt}{20pt}$1^\mathrm{st}$ & $-32.562$ & 
\shortstack{$-30.889$\\ (+1.673)} & \shortstack{$-29.459$ \\(+3.103)}\\
\rule{0pt}{20pt}$2^\mathrm{nd}$ & $-24.449$ & 
\shortstack{$-24.835$\\ (-0.386)} & \shortstack{$-21.509$ \\(+2.940)}\\
\rule{0pt}{20pt}$3^\mathrm{rd}$ & 2.577  & 
\shortstack{3.329\\ (+0.752)} & 
\shortstack{3.802 \\(+1.225)} \\
\rule{0pt}{20pt}$4^\mathrm{th}$ & 21.872 &  
\shortstack{20.175 \\(-1.697)} & 
\shortstack{19.529 \\(-2.28)}\\
\rule{0pt}{20pt}$5^\mathrm{th}$  & 32.562 & 
\shortstack{30.376 \\(-2.186)} & 
\shortstack{26.522 \\(-6.040)}\\
\hline 
DS (cm$^{-1}$) & 218.75& 228.13 (+9.38) & 205.01 (-13.74) \\
\end{tabular}
\end{ruledtabular}
\caption{\label{tab:error_dl}
Comparison between the exact excitation energies and those calculated by DL-VQD and Post-DL procedure with a noisy simulator.
These energies represent the converged values from Fig.\ \ref{post_dl_dl_vqd_result}, with the corresponding absolute errors shown in parentheses.
DS refers to the Davydov splitting, which is the energy difference between the $2^\mathrm{nd}$ and $3^\mathrm{rd}$ excitations.}
\end{table}

Next, we perform deep learning (DL) based error mitigation of the final VQD results, denoted as Post-DL, as shown in Fig.\ \ref{DL}(b).
The corresponding converged values are summarized in Table \ref{tab:error_dl}.
The absolute errors are up to 2.186 meV, while the errors for the $3^{\text{rd}}$ and $4^{\text{th}}$ excitations are less than 1 meV.
The Post-DL Davydov splitting is 228.13 cm$^{-1}$, differing by less than 10 cm$^{-1}$ from the exact value.

It has been shown that even in a noisy environment the optimal parameters in the wave function ansatz can still be found with a variational algorithm \cite{sharma2020noise}.
In our work, Post-DL is applied only to the final state after the variational process to mitigate the error of the optimized wave function.
Hence, the Post-DL expectation values match the classical result in the noiseless limit.

As DL-based mitigation can also be performed during the variational process, it is also worthwhile to benchmark the effectiveness of different DL-based mitigation schemes.
We compare Post-DL results to simulations that perform DL-based error mitigation during each VQD iteration, denoted as DL-VQD, as shown in Fig.\ \ref{post_dl_dl_vqd_result}.
The corresponding converged values are summarized in Table \ref{tab:error_dl}.
The absolute errors are up to 6.040 meV, which are higher compared to the Post-DL results.
Mitigating the measured results during the variational algorithm does not benefit the parameter optimization in the circuits, as we conclude from the DL-VQD expectation values being further from the classical results than Post-DL data (see Table \ref{tab:error_dl}).
Moreover, when performing deep learning predictions during each run of the variational process, the time required to utilize the quantum machine is longer than with Post-DL.
Additionally, instead of predicting a few measured results at a time during optimization with DL-VQD, batch prediction can be performed to predict all the results at once for Post-DL.
Hence, Post-DL outperforms DL-VQD in terms of both accuracy and computational resources.
After validating the feasibility and effectiveness of the Post-DL technique on the Qiskit simulator, we next perform the same procedure on the $ibmq\_jakarta$ device.

\subsection{Quantum Hardware Results}

We use the $ibmq\_jakarta$ device to evaluate the lowest three eigenvalues of the Hamiltonian, Eq.\ \eqref{eq:hamiltonian}, to compute the Davydov splitting.
The converge behavior during the VQD procedure and the absolute error of each eigenstates are shown in Fig.\ \ref{real_hardware_result}. 
To perform Post-DL for these noisy results, we train the FNN using measurements from the quantum hardware and the procedure described in Sec.\ \ref{sec:fnntrain} and Fig.\ \ref{DL}(a).
We use 1050 different datasets, each measured 1024 times, as the entire training dataset and the noiseless data as described in Appendix \ref{appendix_1}.
After training the FNN, we use it to mitigate the error in the noisy results and predict the ideal wave function for evaluating expectation value as shown in Fig.\ \ref{DL}(b). 
In practice, real hardware can be calibrated multiple times a day, which can alter the noise patterns and degrade the performance of the FNN or require retraining.

\begin{table}[tbh]
\begin{ruledtabular}
\begin{tabular}{cccc}
\textrm{Excitation} & \textrm{Noisy (meV)}& \textrm{PS (meV)}& \textrm{Post-DL (meV)} \\
\hline
\rule{0pt}{20pt}$1^\mathrm{st}$ &  \shortstack{$-24.196$\\ (+8.366)}  & \shortstack{$-27.013$ \\(+5.549)}& \shortstack{$-28.941$\\ (+3.621)}\\
\rule{0pt}{20pt}$2^\mathrm{nd}$ &  \shortstack{$-18.876$ \\(+5.573)}& \shortstack{$-21.043$ \\(+3.406)}& \shortstack{$-23.047$\\ (+1.402) }\\
\rule{0pt}{20pt}$3^\mathrm{rd}$ & \shortstack{$2.381$\\ (-0.196)} & \shortstack{ 2.665 \\(+0.088)}& 
\shortstack{2.801\\ (+0.224)} \\
\hline 
DS (cm$^{-1}$) & 172.18 (-46.57) & 192.03  (-26.72) & 209.38  (-9.37)\\
\end{tabular}
\end{ruledtabular}
\caption{\label{tab:real_error_dl}
Comparison between the exact excitation energies and those calculated by performing post selection and deep-learning based mitigation after optimization with $ibmq\_jakarta$ quantum hardware.
These energies represent the converged values from Fig.\ \ref{real_hardware_result} with the corresponding absolute errors shown in parentheses.
DS refers to the Davydov splitting, which is the energy difference between the $2^\mathrm{nd}$ and $3^\mathrm{rd}$ excitations.}
\end{table}

On real quantum hardware, the Post-DL scheme mitigates the error from 46.57 meV to 9.37 meV, similar to the improvement for simulated noise.
The corresponding Davydov splitting differs by less than 10 cm$^{-1}$, as summarized in Table \ref{tab:error_dl}.
Notably, the absolute error for the mitigated results with post-selection is generally larger than that with Post-DL (see Table \ref{tab:real_error_dl}), as it only follows the particle number conservation and does not explicitly capture the relationship between noiseless and noisy data.

While the Post-DL technique achieves an accuracy of about 5 meV, which is sufficient to compare Davydov splitting to experiment, this technique can be used in conjunction with other error mitigation methods to further reduce error.
It is expected that reducing the level of noise in the training data set can improve the performance of deep learning techniques, as the model will have less non-linear relationship to capture. 
This can be achieved through error correction techniques, such as applying dynamical decoupling to idle qubits during simulation, which significantly reduces incoherent errors \cite{viola1999dynamical}. 
Additionally, the Pauli twirling technique can mitigate coherent errors \cite{bennett1996mixed}, which are more difficult to address by deep-learning technique compared to incoherent errors \cite{zhukov2024quantum}.
This can be implemented by adding a decoupling sequence into the idle qubit in the ladder structure of our ansatz in Fig.\ \ref{w_state_vqd} and twirling the CNOT gate with random Pauli gates.
\section{Conclusions}

In this work, we first utilized variational quantum deflation to solve the eigenstates of the Frenkel Hamiltonian using our proposed ansatz for describing the wave function of the Frenkel exciton, which enables efficient evaluation of overlap integrals.
We validated the process by obtaining consistent results with those from exact diagonalization. 
Using a noise model based on real hardware, we performed noisy simulations and found that post-selection can mitigate the error from 42.57 cm$^{-1}$ to 10.81 cm$^{-1}$. 
To mitigate errors, we developed a deep-learning-based mitigation scheme that further reduces the absolute error of observables compared to post-selection technique.
Finally, our calculations on the $ibmq\_jakarta$ machine obtained observables with an absolute error of 9.37 cm$^{-1}$ compared to 26.72 cm$^{-1}$ with the post-selection technique.
As the error is estimated to approach 1 meV (8.06 cm$^{-1}$), comparable to the resolution of optical spectra, our results illustrate usefulness even on early quantum hardware.
As the circuit depth scales linearly with system size, it is expected that our method can handle systems with over 100 molecules on current devices.
However, training the neural network to learn high-dimensional wavefunctions may require meticulous care, as the training cost also increases.
To further address the uncertainty of noise patterns caused by calibration, we anticipate that utilizing dynamical decoupling and Pauli twirling in conjunction with our mitigation scheme can further reduce the error and enhance the effectiveness of the deep-learning-based mitigation.

\appendix
\section{\label{appendix_1}wavefunction}
For the wave function shown in Fig. \ref{w_state_vqd}, the $n$ parameter are used to determine the coefficient for a total of $n+1$ bases.
The first rotation-y gate is used to determine the coefficient for the first basis in ascending order, which is $|0...0001\rangle$, by the following equation:
\begin{equation}
\label{P1}
C_1 = \cos{(P_1/2)}
\end{equation}
where $C_1$ denotes the coefficient of the first basis and $P_1$ refers to the first parameter from the left-hand side of the wave function.
The $\sin{(P_1/2)}$ part will be further divided and distributed to the rest of the bases by the two-qubit gate.
The two-qubit gate splits the amplitude by $\sin{(P)}$ and $\cos{(P)}$ using a given parameter $P$ when the controlled qubit is $|1\rangle$ (which is our case).
In our wave function, we take the sine part from the two-qubit gate as the coefficient of the basis and pass and divide the cosine part for the remaining bases.
For example, $C_2$ and $C_3$ can be described by the following equation.
\begin{equation}
\label{P1}
C_2 = \sin{(P_1/2)}\sin{(P_2)}, \\
C_3 = \sin{(P_1/2)}\cos{(P_2)}\sin{(P_3)}
\end{equation}
where $\sin{(P_1/2)}$ is the remaining amplitude from evaluating $C_1$ and $\cos{(P_2/2)}$ is the remaining amplitude from evaluating $C_2$.
Generally, one can write down the relationship between the n parameters and the corresponding coefficients for the $3^{rd}$ to $n^{th}$ bases as follows:
\begin{equation}
C_i = \sin{\frac{P_1}{2}} \prod_{j=2}^{i-1} \sin{P_j} \cos{P_i}, \ 2<i<n+1
\end{equation}
as for the last term, it will be the product of all sine terms:
\begin{equation}
C_{n+1} = \sin{\frac{P_1}{2}} \prod_{j=2}^{i} \sin{P_j}
\end{equation}

\bibliographystyle{apsrev}
\bibliography{main.bib}

\beginsupplement

\newpage

\begin{center}
\textbf{
\large Quantum Computing for Frenkel Hamiltonian and Error Mitigation with Deep
Learning \\\vspace{0.3 cm}}

Yi-Ting Lee $^{1}$, Vijaya Begum-Hudde$^{1}$, Barbara A.\ Jones$^{2}$, Andr\'e Schleife$^{1,3,4}$

\small

$^1$\textit{Department of Materials Science and Engineering, University of Illinois at Urbana-Champaign, Urbana, IL 61801, USA}

$^2$\textit{IBM Research Almaden Lab, 650 Harry Rd, San Jose, CA 95120}

$^3$\textit{National Center for Supercomputing Applications, University of Illinois at Urbana-Champaign, Urbana, IL 61801, USA}

$^4$\textit{Materials Research Laboratory, University of Illinois at Urbana-Champaign, Urbana, IL 61801, USA}
\end{center}

\begin{figure}[h]
\includegraphics[width=0.63\columnwidth]{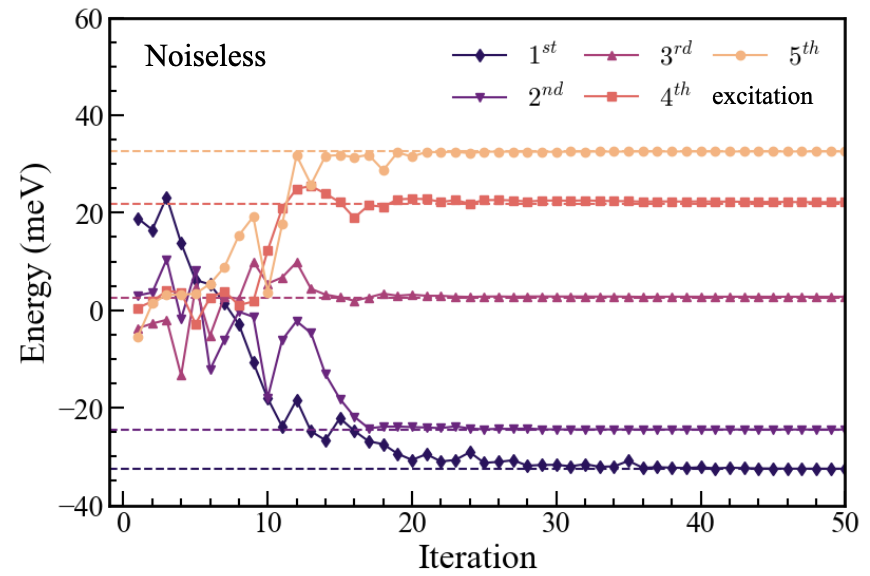}
\caption{\label{noiseless} Noiseless simulations of Frenkel Hamiltonian of single layer anthracene comprising 5 molecules. Dashed lines are the reference obtained from exact diagonalization.}
\end{figure}

\begin{table}[h]
\begin{ruledtabular}
\begin{tabular}{cccccccc}
Qubit & \shortstack{T1 \\ (us)} & \shortstack{T2 \\ (us)}  & 
\shortstack{Readout \\ assignment error}&
\shortstack{Prob meas0 \\ prep1} &
\shortstack{Prob meas1 \\ prep0} &
\shortstack{ID/RZ/ \\ X error} &
\shortstack{CX error \\ (pair: error)} \\
\hline
\centering 0 & 120.7 & 179.57 & 0.013 & 0.0194 & 0.0066 & 2.222e-4 
& \shortstack{0-1: 9.953e-03} \\
\hline
1 & 95.05 & 118.05 & 0.012 & 0.0206 & 0.0034 & 4.700e-4 
& \shortstack{ 1-0: 9.953e-03 \\ 1-2: 8.757e-03 \\ 1-4: 7.341e-03} \\
\hline 
2 & 99.19 & 102.37 & 0.0205 & 0.0276 & 0.0134 & 3.033e-4 
& \shortstack{2-1: 8.757e-03 \\ 2-3: 1.521e-02} \\
\hline
3 & 73.33 & 92.97 & 0.0147 & 0.0244 & 0.005 & 2.490e-4 
& \shortstack{3-2: 1.521e-02 \\ 3-5: 1.094e-02} \\
\hline
4 & 124.84 & 124.95 & 0.021 & 0.0314 & 0.0106 & 3.418e-4 
& \shortstack{4-1: 7.341e-03 \\ 4-7: 1.517e-02} \\
\hline
5 & 84.14 & 87.29 & 0.0152 & 0.0238 & 0.0066 & 2.087e-1 
& \shortstack{5-3: 1.094e-02 \\ 5-8: 5.423e-03} \\
\hline
6 & 83.98 & 14.29 & 0.0375 & 0.0678 & 0.0072 & 3.990e-4 
& \shortstack{6-7: 9.300e-03} \\
\hline

7 & 113.13 & 62.98 & 0.0271 & 0.0348 & 0.0194 & 3.193e-4 
& \shortstack{7-4: 1.517e-02 \\ 7-6: 9.300e-03 \\
 7-10: 6.985e-03} \\
\hline
8 & 94.05 & 93.62 & 0.0149 & 0.0224 & 0.0074 & 2.350e-4 
& \shortstack{8-5: 5.423e-03 \\ 8-9: 9.098e-03 \\ 8-11: 7.540e-03} \\
\hline
9 & 95.65 & 85.88 & 0.0127 & 0.0214 & 0.004 & 4.131e-4 
& \shortstack{9-8: 9.098e-03} \\ 
\hline
10 & 100.86 & 101.47 & 0.0097 & 0.0156 & 0.0038 & 2.901e-4 
& \shortstack{10-7: 6.985e-03 \\ 10-12: 1.282e-02} \\
\hline
11 & 12.95 & 23.61 & 0.0187 & 0.0292 & 0.0082 & 4.102e-4 
& \shortstack{11-8: 7.540e-03 \\ 11-14: 1.076e-02} \\
\hline

12 & 109.52 & 137.63 & 0.0183 & 0.028 & 0.0086 & 1.831e-4 
& \shortstack{12-10: 1.282e-02 \\ 12-13: 5.793e-03 \\ 12-15: 4.944e-03}\\
\hline
13 & 88.5 & 109.56 & 0.0141 & 0.0264 & 0.0018 & 1.995e-4
& \shortstack{13-12: 5.793e-03 \\ 13-14: 1.292e-02} \\
\hline

14 & 123.8 & 117.41 & 0.0261 & 0.043 & 0.0092 & 5.657e-4 
& \shortstack{14-11: 1.076e-02 \\ 14-13: 1.292e-02} \\
\hline
15 & 151.76 & 203.02 & 0.0116 & 0.018 & 0.0052 & 1.732e-4 
& \shortstack{15-12: 4.944e-03}\\

\end{tabular}
\end{ruledtabular}
\caption{\label{tab:hardware_parameters}
Parameter used to construct the noise model for performing noisy simulation with Qiskit simulator.
}
\end{table}

\begin{figure}[h]
\includegraphics[width=0.6\columnwidth]{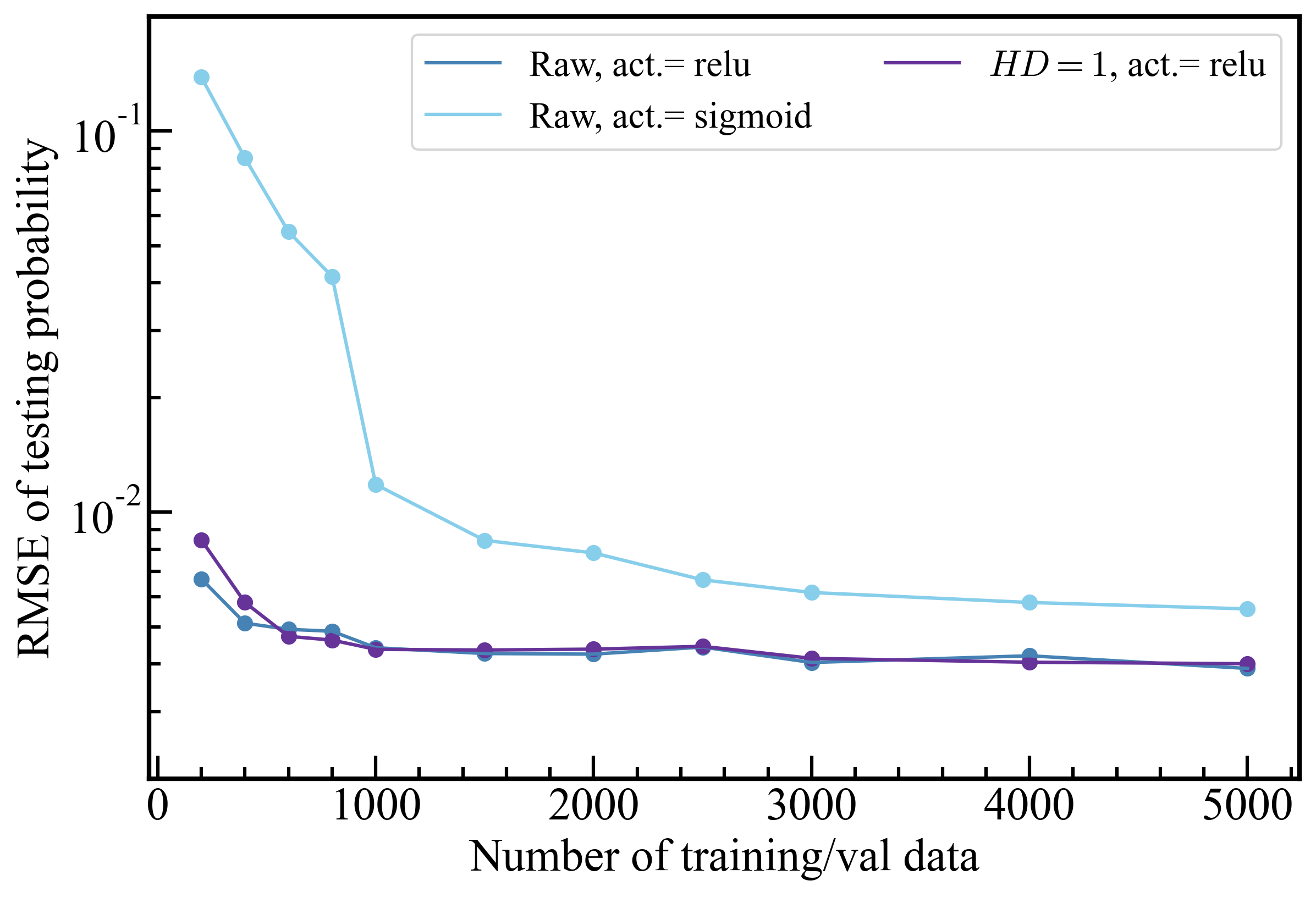}
\caption{\label{dataset_training} Noiseless simulations of Frenkel Hamiltonian of single layer anthracene comprising 5 molecules. Dashed lines are the reference obtained from exact diagonalization.}
\end{figure}

\begin{table}[h]
\begin{ruledtabular}
\begin{tabular}{cccc}
\textrm{batch size}& \textrm{Epoch}& \textrm{Learning rate} & \textrm{Nodes for each layers}\\
\hline
32 & 200 & 1E-4 & 32-32-32-5
\end{tabular}
\end{ruledtabular}
\caption{\label{tab:bayisen}
Optimized Hyperparameters for training feedforward neural networks}
\end{table}

\begin{figure}[h]
\includegraphics[width=0.9\columnwidth]{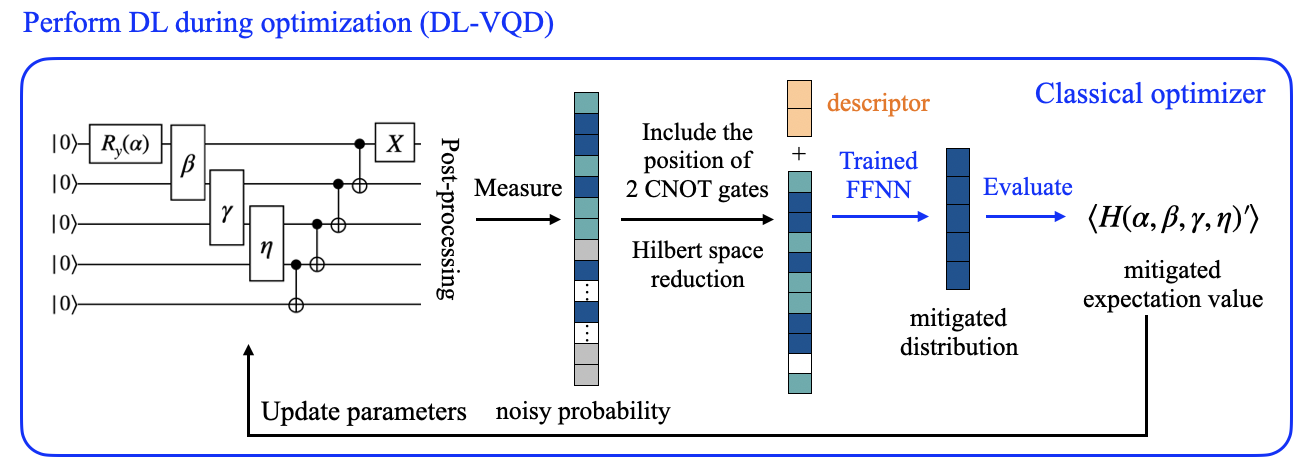}
\caption{\label{DL_VQD} Workflow for applying the deep learning error mitigation technique during the variational process, referred to as DL-VQD.
}
\end{figure}

\begin{figure}[h]
\includegraphics[width=1\columnwidth]{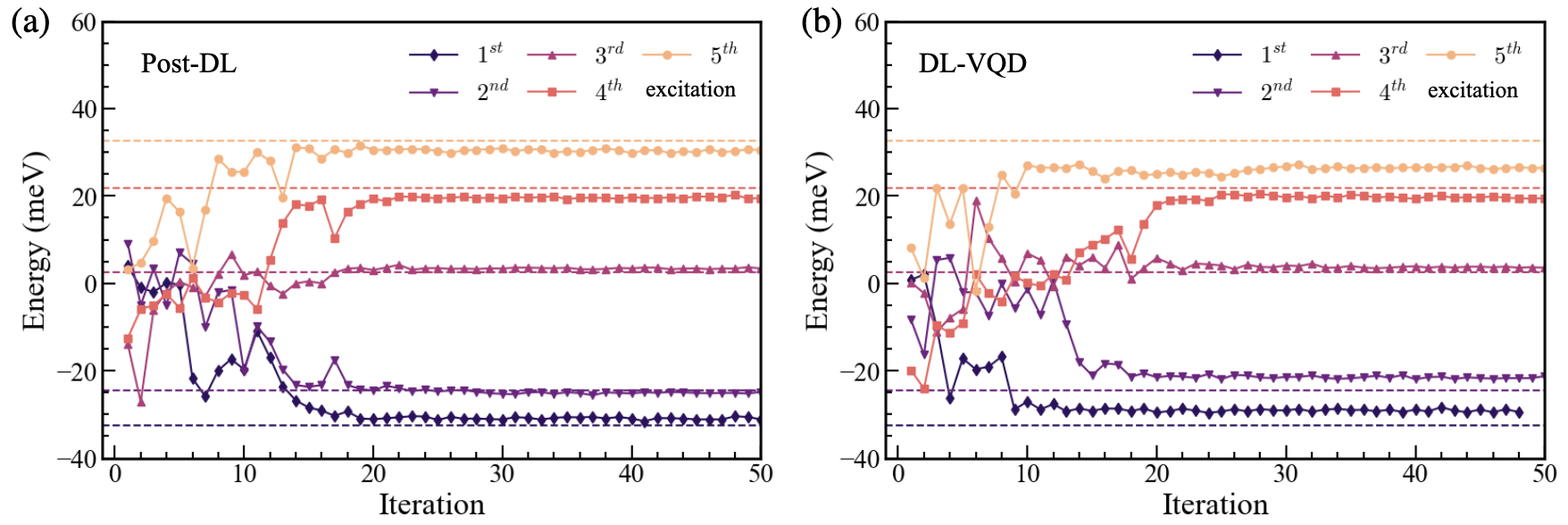}
\caption{\label{post_dl_dl_vqd_result} Noisy simulation results using the Qiskit simulator and the noise model from $ibm_guadalupe$, with (a) Post-DL and (b) DL-VQD. The dashed lines represent the reference obtained from exact diagonalization.
}
\end{figure}

\begin{figure}[h]
\includegraphics[width=1\columnwidth]{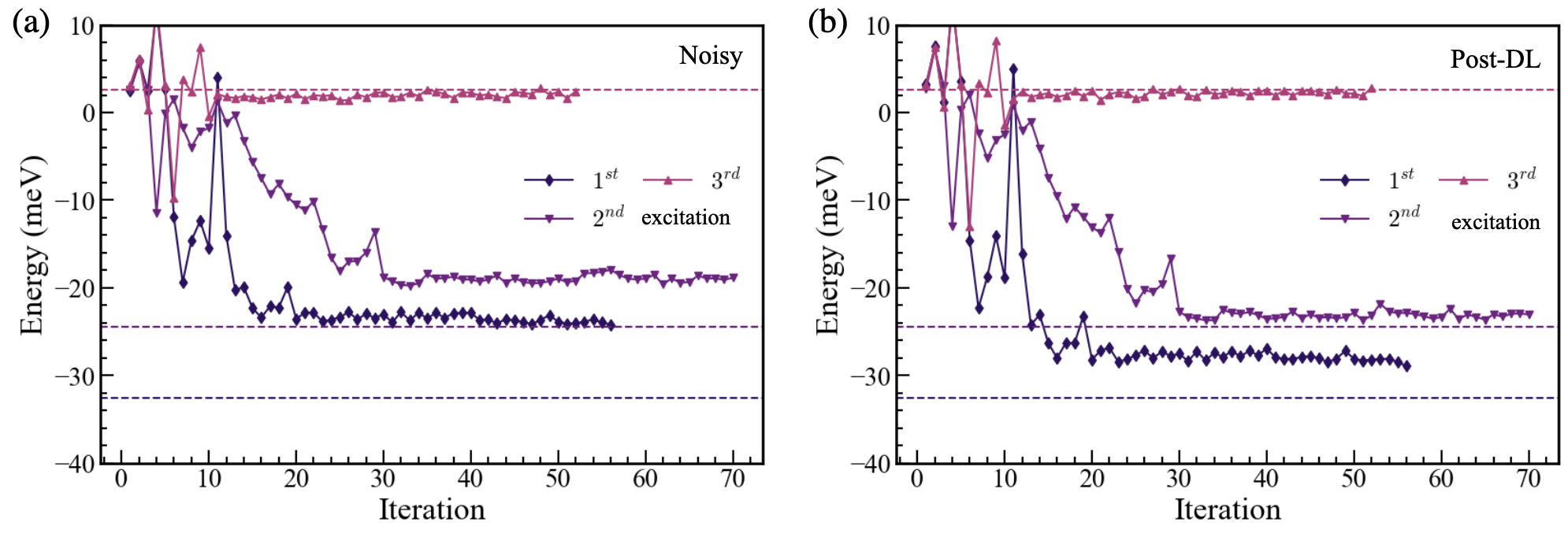}
\caption{\label{real_hardware_result} Quantum simulations of Frenkel Hamiltonian using  $ibmq\_jakarta$ quantum hardware. (a) Raw results without performing any error mitigation. 
(b) Post-DL mitigated results. 
Colored lines are the results under noisy environment and dashed line are the exact ground truth.}

\end{figure}

\end{document}